\documentclass[%
 reprint,
 amsmath,amssymb,
 aps,
]{revtex4-1}

\usepackage{blindtext}
\usepackage{booktabs}
\usepackage{graphicx}% Include figure files
\usepackage{bm}% bold math
\usepackage[utf8]{inputenc}
\usepackage{amsmath,color}
\usepackage{amsfonts}
\usepackage{amssymb}
\usepackage{graphicx}
\usepackage{soul}
\usepackage{ulem}
\usepackage{parskip}
\usepackage{natbib}
\usepackage{algpseudocode}
\usepackage{blindtext}
\usepackage{booktabs}
\usepackage{graphicx}
\usepackage{algcompatible}
\usepackage{multirow}

\begin{document}

\newcommand{\reffig}[1]{Figure \ref{#1}}. % For figure reference! Important! DO NOT delete!

%\preprint{APS/123-QED}

\title{Edge Augmentation on Disconnected Graphs via Eigenvalue Elevation}% Force line breaks with \\%

\author{Tianyi Li}
\thanks{tianyi.li@cuhk.edu.hk}

\affiliation{
$^1$Department of Decision Sciences and Managerial Economics, CUHK Business School, Hong Kong, China\\
$^2$CAS Key Laboratory for Theoretical Physics, Institute of Theoretical Physics, Chinese Academy of Sciences, Beijing 100190, China}


\begin{abstract}
The graph-theoretical task of determining most likely inter-community edges based on disconnected subgraphs' intra-community connectivity is proposed. An algorithm is developed for this edge augmentation task, based on elevating the zero eigenvalues of graph's spectrum. Upper bounds for eigenvalue elevation amplitude and for the corresponding augmented edge density are derived and are authenticated with simulation on random graphs. The algorithm works consistently across synthetic and real networks, yielding desirable performance at connecting graph components. Edge augmentation reverse-engineers graph partition under different community detection methods (Girvan-Newman method, greedy modularity maximization, label propagation, Louvain method, and fluid community), in most cases producing inter-community edges at $>50\%$ frequency. %(105)
\end{abstract}

\maketitle

It is common in social studies employing network analysis that the entire network is comprised of disconnected communities, e.g., villages \citep{BCet2013, CDet2015} or schools \citep{PSet2016} in a geological region, or online chatrooms on a digital platform \citep{H2010}. Analysis is conducted on each distinct community, and results are compared across multiple communities. Although communities possess distinct features, they nevertheless share certain commonalities that are indicative of the background graph consisting of them all. It is useful to recover this background graph from these disconnected components, in order that tasks that operate on the entire graph can be conducted. 

This asks for augmentation of the disconnected graph by introducing new edges. We view the observed subgraphs as downsampled from the entire background graph consisting of all nodes, and try to recover a more detailed connectivity between existing nodes. In particular, complementing existing graph connectivity within subgraphs (i.e., intra-community edges), we establish connections between current communities (i.e., inter-community edges). That is, address the following task: \textit{based on subgraphs' intra-community connectivity, determine most likely inter-community edges that are currently missing on the disconnected graph.}

This graph-theoretical task is seldom touched to the best of our knowledge. More broadly, despite being an important problem, \textit{graph augmentation} has yet to receive substantial research attention. Previous studies formulate graph augmentation as an optimization problem -- determining a minimum-cost set of edges to add to a graph to satisfy a specified property, such as biconnectivity, bridge-connectivity or strong connectivity \citep{FJ1981}. Since the problem is NP-hard, approximate algorithms are developed \citep{FJ1981,WN1987,CS1989,KT1993}. This task witnesses an unexpected revival in recent works \citep{KWet2019,RHet2019,GBet2021,ZLet2021}, where graph augmentation, as a subordinate of data augmentation, is studied to improve the generalizability of machine learning on graphs (see reviews in \cite{DXet2022,ZLet2022}). Those studies focus primarily on edge dropping for removing the over-smoothing of graph neural networks, or on node augmentation for enriching the dataset, paying less attention to edge augmentation.\\
    
\textit{Method.}-- Consider graph $G$ consisting of $M$ subgraphs (disconnected communities): $G(N,E) = \{G_m\}_{m = 1:M}=\{(N_m,E_m)\}_{m = 1:M}$. Subgraph $G_m$ has $|N_m|$ nodes and $|E_m|$ edges: $N_m = \{n^{m,i}\}, E_m = \{e^{m,i}\}$. Subgraphs $G_1, G_2, ..., G_M$ are ordered by their size, e.g., $|N_1|<|N_2|$. 

We consider that the observed edges $\{E_m\}_{m = 1:M}$ are a subset of the edges in the whole graph. With additional edges $\Delta E$, the augmented edge set (\textit{sup}) could facilitate a fully connected (\textit{fc}) graph where current communities are all interconnected, and even beyond, until possibly reaching the complete graph having $|N|(|N|-1)/2$ edges. 

We parameterize edge sets at different scenarios with edge density $\theta \ge 1$: at the current edge set, $\theta=1$; at augmented edge sets, $\theta>1$; $\theta$ may reach maximum $\theta_{max}$, which is no greater than $|N|(|N|-1)/2|E|$ at the complete graph (see below). Edge sets are further distinguished by the multiplicity of zero eigenvalues of $G$'s Laplacian, denoted with $M^{\dagger}$, which equals the number of connected components (i.e., disconnected communities) in the graph \citep{CG1997,V2007}. For current graph $G$, the number of components $M^{\dagger} = M$; with additional edges, $M^{\dagger}$ decreases, until the graph is fully connected, in which case $M^{\dagger} = 1$. The span of edge set scenarios is:
\begin{widetext}
\begin{equation}
E(\theta=1,M^{\dagger}= M) \subseteq E_{sup}(\theta>1,M^{\dagger} \le M) \subseteq E_{fc}(\theta>1, M^{\dagger}=1) \subseteq E_{max}(\theta=\theta_{max} \le \frac{|N|(|N|-1)}{2|E|},M^{\dagger}=1).
\end{equation}
\end{widetext}
If we have information on node attributes, we can use this data to project which edges are likely to appear, for example, considering the problem of link prediction \citep{LK2007,LZ2011}. When we do not have such data and only have graph connectivity information, it is difficult to employ link prediction methods to establish inter-community edges -- as connectivity-based link prediction relies on common neighbors between a node pair to predict their potential linkage, new edges are not connecting nodes having zero existing path. 

\begin{figure}[htp!]
\centering   
	\includegraphics[width=3.5in]{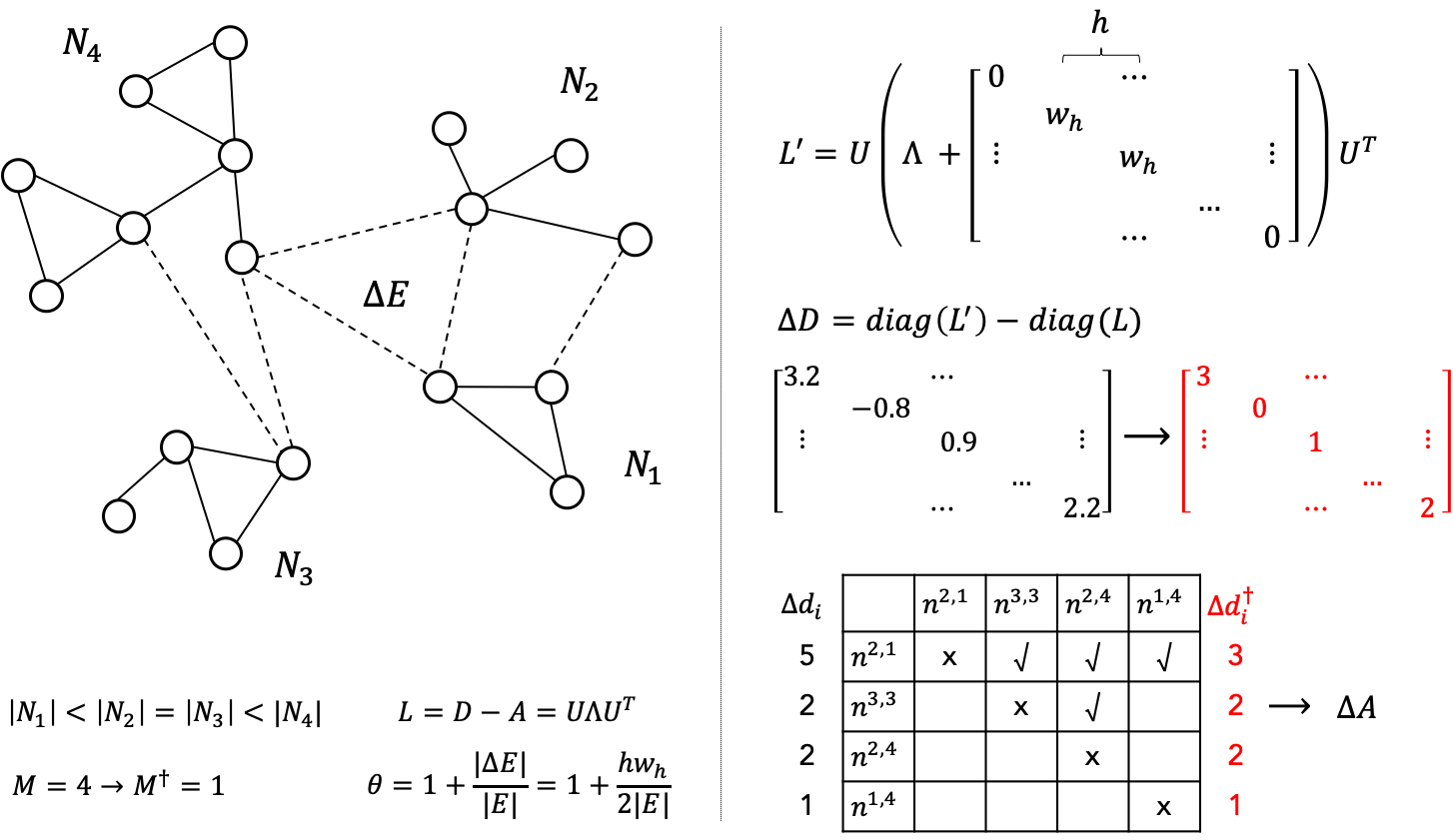}  
\caption{Illustration of graph augmentation via eigenvalue elevation.}
\end{figure}

\textit{(A) Graph augmentation via eigenvalue elevation}

In this case, we augment the graph through working on the spectrum of graph Laplacian. Consider graph $G$'s adjacency matrix $A = \{a_{ij}\}$. Graph Laplacian $L = D - A$ where $D$ is diagonal matrix whose elements are node degrees, $D = diag(d_{i})$. Eigendecompose $L = U\Lambda U^T$, where $\Lambda$ is diagonal matrix whose elements are eigenvalues of $G$, $\Lambda = diag(\lambda_1,\lambda_2, ...,\lambda_N)$, and $U = (v_1,v_2, ..., v_N)^T$ are the corresponding eigenvectors. Arrange eigenvalues in increasing order, with $\lambda_1=0$ being the smallest eigenvalue. As the graph has $M>1$ disconnected communities, the first $M$ eigenvalues $\lambda_1, ..., \lambda_M$ are zero. To connect communities, we elevate zero eigenvalues to reduce the number of connected components. Consider $h\in [1,M-1]$ eigenvalues elevated uniformly to $w_h$ from 0, encoded in $ \Lambda_{\delta} = diag(0,\underbrace{w_h,...,w_h}_{h}, 0,...,0)$. The modified Laplacian 
\begin{equation}
L' = U(\Lambda + \Lambda_{\delta})U^T = L + U\Lambda_{\delta}U^T = L + \sum_{q=1:h} w_h v_q v_q^T,
\end{equation}
where $v_q = (u_{1q}, u_{2q}, ..., u_{Nq})^T$ are the eigenvectors corresponding to the $h$ elevated eigenvalues ($u_{\bullet}$ are elements in $U$). \\

\textit{(B) Upper bound for edge density}

As $tr(v_q v_q^T) = v_q^T v_q = 1$, there is
\begin{equation}
tr(L') = tr(L) + \sum_{q=1:h} w_h = tr(L) + h w_h.
\end{equation}
On the other hand, $tr(L) = tr(D-A) = tr(D) = vol(A) = 2|E|$ \citep{N1}, and similarly, $tr(L') = vol(A') = 2|E'|$, i.e., the trace of $L'$ equals the volume (e.g., twice the edge number) of the augmented graph $A'$. Recall the definition of $\theta$ as the indication of edge density, $|E'| = \theta|E|$, there is the expression
\begin{equation}
\theta = 1 + \frac{h w_h}{2|E|}.
\end{equation}
By determining the number of eigenvalues to be elevated and the amplitude of elevation, $h$ and $w_h$, we control the density $\theta$ of augmented edges. 
We derive the upper bound for $\theta$. For each value of $h$, the number of edges in the augmented graph is the largest if the graph is turned into a complete graph on the largest connected component having $N-(M-1-h)$ nodes, besides the $h$ isolated nodes each forming a single-node component. That gives 
\begin{equation}
\begin{aligned}
&\hat{\theta}^{max}_h = [|N|-(M-1-h)][|N|-(M-1-h)-1]/2|E| \\
&\Longrightarrow \hat{w}_h^{max} = (|N|-M+1+h)(|N|-M+h)-2|E|/h. 
\end{aligned}
\end{equation}
Both $d\hat{w}_h^{max}/dh$ and $d\hat{\theta}^{max}/dh$ are positive, thus
\begin{equation}
\begin{aligned}
&\hat{w}^{max} = \underset{h}{max}\ \hat{w}_h^{max} = |N|(|N|-1)-2|E|/(M-1), \\
&\hat{\theta}^{max} = \underset{h}{max}\ \hat{\theta}_h^{max} = |N|(|N|-1)/2|E|. 
\end{aligned}
\end{equation}
This maximum value of $w_h$ cannot be realised, however. As $w_h$ continues to grow, it will surpass the largest eigenvalue $\lambda_N$ of the original $L$ and become the largest eigenvalue of $L'$. Its value is then capped by the size of the augmented graph \citep{AM1985}. For $h\ge M-1$, the graph has one connected component, thus the largest eigenvalue $w_h$ is less than graph size $|N|$ (in fact, $w_h\le|N|$ for any $h$). For $h<M-1$, graph has more than one component. Consider addition node degree $\Delta d_i$. On the original $L$, existing node degree $d_i = \sum_{K=1}^{|N|}\lambda_k u_{ik}^2$; on $L'$, augmented node degree $d'_i$ takes similar form, thus
\begin{equation}
    \Delta d_i = d'_i - d_i = \sum_{K=1}^{|N|}\lambda'_k u_{ik}^2 - \sum_{K=1}^{|N|}\lambda_k u_{ik}^2 = w_h\sum_{K=2}^{h}\lambda_k u_{ik}^2.
\end{equation}
Consider elements $u_{ik}$ of the same row at different eigenvectors to be roughly independent and identically distributed. As $\sum_i u_{ik}^2 = 1$, $\bar{u_{ik}^2}\sim 1/|N|$; thus $\Delta d_i \sim w_h h/|N|$. Ensuring successful assignment of extra degree $\Delta d_i$, the augmented $d'_i$ is upper bounded by the maximum possible size of the augmented graph, which is adding the size of the largest $h+1$ components among the $M$ components \citep{N2}. This is for a predefined graph topology in terms of the sequence $\{|N_m|_{m=1:M}\}$ of community sizes; considering arbitrary graph topologies, the largest size of the aggregated $h+1$ components is when the rest $M-(h+1)$ components are all isolate nodes. Each degree in $d'_i$ has 1/2 probability to be in $\Delta d_i$ (the other 1/2 probability in $d_i$); thus there is  
\begin{equation}
\begin{aligned}
2\Delta d_i &\le \sum_{s=M-h}^{M-1}|N_s| \le |N| - (M-h-1) - 1 \\
&\Rightarrow 2 \frac{w_h h}{|N|} \le |N| - M + h. \\
\end{aligned}
\end{equation}
Overall, the upper bound for $w_h$ is
\begin{equation}
w_h \le \Biggl\{
\begin{aligned}
&\frac{|N|(|N| - M + h)}{2h},\ \ \ && 0 < h < M-1 \\
&|N|. \ \ \ && \forall h \\
\end{aligned}
\end{equation}
The two upper bounds coincide when $|N|(|N| - M + h)/2h = |N|$ at $h=M-1$, which gives $M = |N|/2$, i.e., the number of connected components reaches half graph size. One such case is Erdos-Renyi (ER) graph at the threshold for giant component emergence \citep{ER1960}, i.e., $G=ER(|N|, p=1/|N|)$, in which case multiple non-trivial components exist besides the giant component, while the rest bulk of components are isolates. Most real-world graphs have $M < |N|/2$, in which case the second upper bound $w_h<|N|$ is tighter, entering into the region of $0 < h < M-1$; extreme graph topology may have $M > |N|/2$, leaving a discontinuity of this upper bound.

Consider $M < |N|/2$ at common graphs, for $\theta$, there is upper bound
\begin{equation}
\theta^{max}_h = 1 + \frac{h w^{max}_h}{2|E|} \le 1+\frac{h|N|}{2|E|} = 1 + \frac{h}{k_{ave}},
\end{equation}
where $k_{ave}$ is average degree of the original graph $G$. Further, $\theta^{max} = 1 + \frac{(M-1) w^{max}}{2|E|} \le 1+\frac{(M-1)|N|}{2|E|} = 1 + \frac{M-1}{k_{ave}}$, considering all values of $h$. As $|N|$ is always no greater than $\hat{w}_h^{max} =  [(|N|-M+1+h)(|N|-M+h)-2|E|]/h$ (equation (5)), $\hat{w}_h^{max}$ will be realized only in one extreme case: $h=M-1$ and the connected component is complete graph. In general, the above upper bound $\theta_h^{max}$ is tighter than $\hat{\theta}^{max} = |N|(|N-1|)/2|E|\sim |N|/k_{ave}$, as in (6). (10) suggests that the amplitude of successful graph augmentation depends on original graph topology, via $k_{ave}$ and $M$, i.e., depending on original graph's capacity of accommodating new edges. For dense graphs where $k_{ave}$ approaches $|N|$, this upper bound $1+(M-1)/k_{ave}$ approaches $|N|/k_{ave}\sim 1$, converging to the complete graph scenario at (6).\\

\textit{(C) Determining new edges}

The location of new edges is determined after we obtain $L'$. Recall $L=D-A$ and $L'=D'-A'$. $D$ and $D'$ are the diagonal elements of $L$ and $L'$, respectively; diagonals of $A$ and $A'$ are zero. We first get $D'$ from the diagonal of $L'$, then indicate the number of additional degrees of each node, $\Delta D$, by comparing $D'$ to $D$. Diagonal elements of $\Delta D$, $\Delta d_i$, may not always be non-negative integer; let $\Delta d_i\leftarrow MAX(\lfloor \Delta d_i \rfloor,0)$ in these cases. The above threshold for $w_h$ ensures that the maximum element in $\Delta D$, i.e., the maximum additional degree of a node, \textit{can} be assigned in the augmented graph. That is, the maximum additional degree is at most the size of the new (largest) connected component minus one; $w_h$ beyond the threshold will lead to unrealistic elements in $\Delta D$ that are too large to be realisable node degrees.

Even below the threshold of $w_h$, the additional node degrees are not guaranteed to be realizable. This is because the set of nodes that are going to have additional degrees, i.e., the nonzero elements of $\Delta D$, may not be as large as some elements of $\Delta D$. Although elements of $\Delta D$ are ensured to be no larger than the component size limit, new degrees are formulated only between nodes corresponding to the nonzero elements of $\Delta D$, not between a node in this set and a node that has zero entry in $\Delta D$. This suggests that realized degrees are no greater than the sum of the diagonals of $\Delta D$, which equals $tr(L')-tr(L) = h w_h$ \citep{N3}. Consequently, the realized number of components, $M^{\dagger}$, can be larger than $M-h$.

We use the following algorithm to realize additional node degrees in $\Delta D$, i.e., to determine the additional adjacency $\Delta A = A'-A$. Sort the nodes having nonzero $\Delta d$ in descending order of $\Delta d$. Starting from the first node in the list, add an edge between the current node and each node below it in the list that still has quota for new degree, if such an edge does not exist; stop if reaching list end or when $\Delta d$ has been all assigned; the realized $\Delta d^{\dagger}$ is then no more than $\Delta d$. Continue until reaching list end. The complete algorithm of graph augmentation via eigenvalue elevation is summarized in the Supplemental Material. As mentioned, truncating additional degrees $\Delta d$ as integers and assigning them only among modified nodes make realized new edges less than projected new edges; edge realization ratio $\phi = 2 \sum \Delta d^{\dagger} /h w_h < 1$.\\

\begin{figure*}[htp!]
\centering   
	\includegraphics[width=7in]{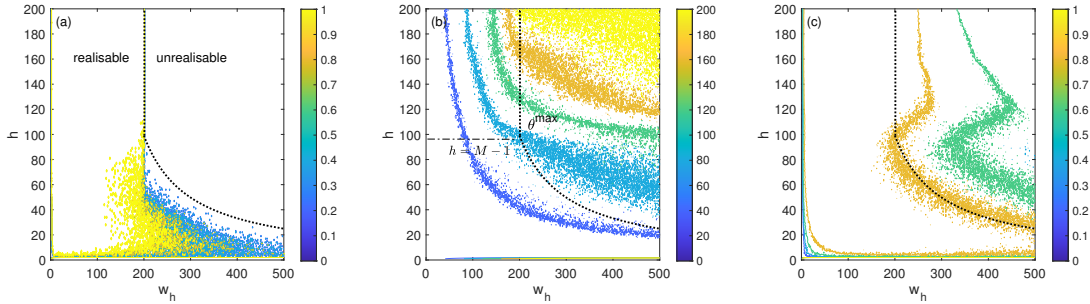}  
\caption{Upper bound for (a) $w_h$, (b) $\theta$; and (c) edge realization ratio $\phi$, in the $(w_h, h)$ space. Dashed lines are (9). Cross point on (b) is $\theta^{max}$ at $h=M-1$. Each point is averaging over 10 Erdos-Renyi random graph instances $ER(|N|=200, p=1/|N|)$.}
\end{figure*}

\textit{(D) Evaluation of augmentation outcome}

The outcome of edge augmentation is evaluated from two aspects. We first check the fraction of inter-community edges among proposed new edges, which include both intra- and inter-community edges. This fraction $\rho$ indicates algorithm's shear ability in establishing inter-community connections. Note that, although in current problem setting (i.e., connecting communities), inter-community edges are preferred over intra-community edges, a large $\rho$ may not be desired in practice, as too much inter-community linkage destroys original community structures. So happens often with large $w_h$: when $w_h$ exceeds original eigenvalues and departs from the bulk of them, it is imposing new low-dimension structures that aggregates existing communities (e.g., \citep{CGet2009}), drastically altering graph topology. In practice, $w_h$ can be chosen at $k_{ave}$ or $\lambda_{ave}$ to prevent this failure.

This ambiguity at $\rho$ brings the second evaluation, where we investigate inter-community edge determination against ground truth, i.e. to which extent the algorithm can recover real inter-community links that are hidden from view. This uses graph data that have ground truth community structures; inter-community edges are covered and are checked against algorithm-determined edges.

In a further sense, as hinted above, since community structure on graphs is ill-defined and there is in fact no ``ground truth'' community \citep{PLet2017,LZ2020}, this task of inter-community edge augmentation can be viewed as the inverse task for community detection. The above edge augmentation algorithm thus reverse-engineers community detection algorithms. Hence the problem of whether the algorithm can recover ground truth inter-community connections is better formulated as to which extent this algorithm can recover the inter-community connections determined by which community detection algorithm, i.e., this edge augmentation heuristic is an effective reverse-engineering companion of which community detection heuristic.\\

\textit{Results.}--
\textit{(A) Upper bound for $w_h$ and $\theta$.} Consider Erdos-Renyi random graph $G=ER(|N|=200, p=1/|N|)$ at the threshold for giant component emergence, in order that multiple non-trivial components exist and that $M = |N|/2$. For each $h\in[0, 200]$ and $w_h\in[0, 500]$, average results over 10 random graphs. Count the number of times $\Delta d^{max}$, determined under $w_h$ and $h$, can be realised during edge augmentation, i.e, when the augmenting node set $|N^{\Delta}| \ge \Delta d^{max}$. Compute realized edge density $\theta = 1 + \sum \Delta d^{\dagger}/|E|$ at each $(w_h, h)$.

Upper bound for $w_h$ holds consistently (Figure 2a). At region $h<M-1$, fewer instances of random ER graphs approach the upper bound, which is for arbitrary graph topology and is bounding at the most extreme case with a giant complete component and $M-h$ isolates. Average component number $M = |N|/2 = 100$; corresponding upper bound for $\theta$, $1+h/k_{ave}$, holds at each value of $h$ and in particular at $h=M-1$ (Figure 2b; $k_{ave}=1$). Consistently, edge realization ratio $\phi$ remains high in the realizable $(w_h, h)$ region, slightly below 1 due to finite effects (Figure 2c) \citep{N4}.

\begin{figure}[h]
\centering   
	\includegraphics[width=2.7in]{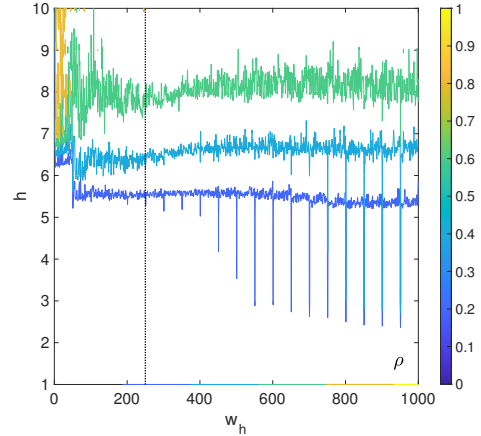}  
\caption{Fraction $\rho$ of inter-community edges among realized edges during the recovery of inter-community connections at SBM. Dashed line marks the boundary of realisable/unrealisable region.}
\end{figure}

\textit{(B) Recovering inter-community connections of SBM.} We evaluate algorithm's ability of recovering inter-community connections at the stochastic block model (SBM) \citep{HLet1983,KN2011}. Initiate $M=5$ blocks each containing 50 nodes, totaling at $|N|=250$; $p_{in}, p_{out} = 0.5, 0.1$; expected inter-block/intra-block edge ratio is 0.8. Hide all inter-block edges and conduct edge augmentation. Average results over 5 random graph instances for each $h\in[1, 10]$ and $w_h\in[0, 1000]$. For this graph, as $|N|(|N| - M + h)/2h$ is always greater than $|N|$, the upper bound $w_h\le |N|$ holds across the range of $h$.

The fraction $\rho$ of inter-block edges among realized edges increases as $h$ increases and roughly remains constant as $w_h$ increases (Figure 3). Inter-block connections become denser when a larger graph eigenspace is being modulated via altering more eigenvalues; the amplitude of augmentation plays a lesser role. With nodes in SBM communities being homogeneous, the recovery rate $\epsilon$ of exact hidden inter-block edges is small ($<0.1$, not shown).

\begin{table*}
\begin{tabular}{c|c|c|c|c|c|c}
\hline
$M^{\dagger}$/$\rho$/$\epsilon$ & $h$ & Girvan-Newman & Greedy modularity & Label propagation & Louvain & Fluid community \\
\hline
Davis southern women & 0 & 2/-/- & 3/-/- & 2/-/-         & 3/-/- & 2/-/- \\
($|N|=32, |E|=89$) & 1  & 2/0/0 & 3/0/0 & 2/0/0          & 3/0/0 & 2/0/0\\
                   & 2  & 2/0/0 & 3/0/0 & 1/0.97/0.25    & 2/0.86/0 & 1/0.65/0.03 \\
                   & 3  & 2/0/0 & 1/0.46/0.11 & 1/0.64/0 & 2/0.97/0.04 & 1/0.49/0.09 \\
                   & 4  & 1/0.44/0.20 & 1/0.74/0.18 & 1/0.48/0 & 1/0.96/0.04 & 1/0.61/0.17 \\
Karate club        & 0  & 2/-/- & 3/-/-       & 4/-/-       & 4/-/-    & 2/-/- \\
($|N|=34, |E|=78$) & 1  & 2/0/0 & 2/0.82/0.11 & 3/1.00/0.06 & 4/0/0    & 2/0/0 \\
                   & 2  & 2/0/0 & 2/0.76/0.21 & 2/0.97/0.12 & 3/0.82/0 & 2/0/0\\
                   & 3  & 2/0/0 & 2/0.55/0.16 & 1/0.66/0.31 & 2/0.74/0 & 1/0.42/0.09 \\
                   & 4  & 1/0.51/0.10 & 2/0.59/0.16 & 1/0.71/0.31 & 1/0.78/0.05 & 1/0.53/0.09 \\
Dolphin             & 0  & 2/-/-  & 4/-/- & 5/-/- & 5/-/- & 2/-/- \\
($|N|=62, |E|=159$) & 1 & 1/0.55/0 & 3/0.06/0 & 4/0.93/0.06 & 4/0.07/0 & 2/0/0 \\
                    & 2 & 1/0.48/0 & 2/0.49/0 & 3/0.77/0.06 & 3/0.58/0 & 1/0.52/0.29 \\
                    & 3 & 1/0.63/0 & 2/0.81/0.14 & 2/0.82/0.12 & 2/0.62/0.03 & 1/0.58/0.24 \\
                    & 4 & 1/0.53/0 & 2/0.75/0.14 & 2/0.87/0.15 & 1/0.67/0.05 & 1/0.53/0.12 \\
Facebook TV show        & 0  & 2/-/- & 59/-/- & 410/0/0 & 49/-/- & 9/-/- \\
($|N|=3892, |E|=17239$) & 1  & 2/0/0    & 27/0.51/0 & 263/0.88/0   & 22/0.48/0 & 6/0.99/0\\
                        & 2  & 1/0.68/0 & 16/0.66/0 & 186/0.80/0   & 11/0.45/0 & 5/0.54/0\\
                        & 3  & 1/0.51/0 & 10/0.63/0 & 94/0.81/0    & 7/0.52/0 & 4/0.63/0\\
                        & 4  & 1/0.60/0 & 6/0.65/0  & 52/0.83/0.01 & 4/0.51/0 & 3/0.80/0 \\
                        & 5  & 1/0.49/0 & 4/0.67/0  & 21/0.85/0.01 & 2/0.46/0 & 3/0.97/0\\
                        & 6  & 1/0.37/0 & 3/0.66/0  & 14/0.86/0.01 & 2/0.51/0 & 2/0.98/0\\
\hline
\end{tabular}%
\caption{Reverse-engineering community detection on real networks via edge augmentation.}
\end{table*}

\textit{(C) Reverse-engineering community detection at real networks.} Apply a panel of community detection algorithms (Girvan-Newman method \citep{GN2002}, greedy modularity maximization \citep{CNet2004}, label propagation \citep{RAet2007}, Louvain method \citep{BGet2008}, and fluid community \citep{PGet2017}) on small-to-medium real networks (Davis southern women \citep{DGet2009}, Karate club \citep{Z1977}, Dolphin network \citep{LSet2003}, and a recent Facebook network \citep{RDet2019}) for demonstration.

On each graph, after community detection, hide inter-community edges and recover via edge augmentation. Set $w_h$ equal to network size; vary $h$. Compute the resulting component number $M^{\dagger}$, fraction $\rho$ of inter-community edges among new edges, and recovery rate $\epsilon$ of hidden ground-truth inter-community edges.

Results are shown in Table 1 (showing one set of results at indeterministic community detection; results are robust across multiple runs). Consistently, for the five community detection methods, as $h$ increases, components are always better connected (i.e., $M^{\dagger}$ decreases; showing the first pair of communities at Girvan-Newman to illustrate $M^{\dagger} = 2 \rightarrow 1$). A small $h$ is sufficient for considerably connecting the partitioned real network (Facebook TV show). In most cases, over half of proposed edges are inter-community ($\rho > 0.5$), while the recovery rate $\epsilon$ of exact ground-truth inter-community connections is not high. Overall, this edge augmentation method most reverse-engineers community detection under label propagation, yielding large $\rho$ across different graphs and $h$ values.\\

\textit{Concluding Remarks.}-- We propose the task on graphs of determining most likely inter-community edges based on components' intra-community connectivity. We develop an algorithm for this edge augmentation task based on elevating zero eigenvalues of graph's spectrum. Upper bounds for eigenvalue elevation amplitude and for the corresponding augmented edge density are derived and are validated on random graphs. Algorithm performs consistently on synthetic and real networks, showing desirable performance at connecting graph components and varied reverse-engineering compatibility towards different community detection methods.

We assume uniform $w_h$ in eigenvalue elevation; non-uniform $w_h$ may lead to more general results. The method can further extend to using alternative graph Laplacian (e.g., normalized, random-walk). The matching of inter-community edge augmentation heuristic with community detection ideas is prefatory and asks extensive empirical analysis.

The ``most likely'' inter-community edges, as we set out to determine, can be defined in different ways. This suggests that edge augmentation outcome can be evaluated on resulting graph's performance at specific tasks that welcomes global connectivity. One suitable task is semi-supervised learning on graphs, for example, using graph convolution neural networks \citep{ZTet2019}, with the convolution taking place in the spectral (e.g., GCN \citep{KW2016}) or spatial space (e.g., DCNN \citep{AT2016}). We investigate which edges to add to the disconnected graph, such that semi-supervised classification can better generalize; that is, the evaluation of inter-community connections anchors on their effect in enhancing classification performance. Application of this edge augmentation algorithm to semi-supervised classification on disconnected graphs points to an interesting future work.


\begin{thebibliography}{26}
\bibitem[{\textit{Note 1}}]{N1}
Assume no self-loops on nodes; diagonal elements of $A$ are zero. Moreover, $tr(L) = tr(\Lambda) = \sum_i \lambda_i$.
\bibitem[{\textit{Note 2}}]{N2}
Removing $h$ zero eigenvalues corresponds to reducing $h$ components and thus aggregating $h+1$ components.
\bibitem[{\textit{Note 3}}]{N3}
For a valid $\Delta D$, let the product $h w_h$ be an odd number when determining $h$ and $w_h$.
\bibitem[{\textit{Note 4}}]{N4}
In the experiment, edge assignment is allowed to proceed in the unrealizable $(w_h, h)$ region, realizing a portion of proposed new degrees. Smaller realization ratio $\phi$ scatters the contour of $\theta$ in this region (Figure 2b).
\end{thebibliography}
\end{document}